\documentclass[sigconf, 10pt]{acmart2}

\usepackage[english]{babel}
\usepackage{blindtext}

\renewcommand\footnotetextcopyrightpermission[1]{} 
\setcopyright{none}

\setcopyright{none}
\newcommand{\PaperTitle}{
Leveraging eBPF and AI for Ransomware Nose Out
}

\newcommand{\hide}[1] {}
\newcommand\sknote[1]{{\color{magenta}[SGK: #1]}}
\renewcommand\sknote[1]{}

\begin{document}
\setlength{\belowcaptionskip}{-6pt}

\title{\PaperTitle}


\author{Arjun Sekar$^\dag$, Sameer G. Kulkarni$^\dag$ and Joy Kuri$^\ddagger$ \\
$^\dag$Indian Institute of Technology Gandhinagar,
$^\ddagger$Indian Institute of Science, India
}


\renewcommand{\shortauthors}{Sekar A., Kulkarni S. G., and Kuri J.}
\acmConference[Arxiv]{Arxiv}{June 2024}{}
\begin{abstract}
\hide{
In this work, we propose 
a two-phased 
approach to detect and deter ransomwares in real-time.  
Towards this objective, we harness the power of eBPF (Extended Berkeley Packet Filter) and machine learning to develop proactive and reactive methods. 
}
In this work, we propose a two-phased approach for real-time detection and deterrence of ransomware. 
To achieve this, we leverage the capabilities of eBPF (Extended Berkeley Packet Filter) and artificial intelligence to develop both proactive and reactive methods.
\hide{
The secondary layer initiates a behavior-based technique focused on monitoring the creation of ransom notes, a distinctive indicator of ransomware activity, through custom-developed eBPF kernel-level tracing programs. 
}
In the first phase, we utilize signature-based detection,
where we employ custom eBPF programs to trace the execution of new processes and perform hash-based analysis against a known ransomware dataset.
In the second, we employ a behavior-based technique that focuses on monitoring the process activities using a custom eBPF program and the creation of ransom notes \textemdash{} a prominent indicator of ransomware activity through the use of Natural Language Processing (NLP). 
By leveraging eBPF's low-level tracing capabilities and integrating NLP based machine learning algorithms, our solution achieves an impressive 99.76\% accuracy in identifying ransomware incidents within a few seconds on the onset of zero-day attacks. 
\end{abstract}

\maketitle
\thispagestyle{empty}

\vspace{-6mm}\section{Introduction}\label{sec:intro}
\hide{Ransomware is a type of malware that takes control of a computer system and encrypts its files, effectively rendering them inaccessible to the victim. The attackers then demand a ransom payment in exchange for the decryption key.
}
Ransomware has emerged as a highly destructive and pervasive form of malware, evolving into the preferred weapon for cybercriminals in recent years \cite{MANSFIELDDEVINE201715}.
As per the Sonicwall Cybersecurity report~\cite{sonicwall_cyber_threat_report_2024}, in 2023 alone, businesses faced a staggering escalation in ransomware attacks, with over 318 million incidents reported - 
roughly an attempt to attack every 0.1 seconds. 
The threat extends beyond individual finances, reaching hospitals, power grids, and entire industries, posing a severe risk to public safety and national security.
The severity of threat posed by ransomwares is highlighted by its reclassification as a top security concern in the latest White House National Cybersecurity Strategy~\cite{national_cybersecurity_strategy_2023}. 
\hide{
This emphasizes the need for a comprehensive federal approach and international collaboration to counteract ransomwares.
The financial toll of ransomware attacks is staggering, with the average cost reaching \$5.3 million in 2023, marking a significant 13\% increase from the previous year’s average of \$4.54 million \cite{threatlabz_ransomware_report_2023}.
It has been estimated that the total loss inflicted by ransomware in 2023 accounts for about \$30 billion. 
Looking ahead, predictions indicate that by 2031, ransomware is poised to inflict a colossal \$265 billion in damages on its victims, driven by a projected 30\% year-over-year growth over the next decade \cite{Morgan_2023}. 
These costs encompass ransom payments, data damage and destruction, lost productivity, intellectual property theft, personal and financial data compromise, post-attack disruptions, forensic investigations, data and system restoration, and reputation harm.

The above statistics advocate the need for developing more efficient ransomware detection and evasion techniques to put an end to its rampage.
}

The most common approach followed by anti-virus solutions includes signature-based detection which struggles to detect zero-day attacks~\cite{ianti_virus}.
\sknote{Discuss pros and cons of the existing approach then lead to how eBPF can help.}
Extended Berkeley Packet Filter (eBPF) strikes the right balance to run lightweight programs securely in kernel space without the need to modify the kernel, and facilitates efficient mode of kernel level observability to implement simple security measures with minimal processing overheads when compared to the solutions developed as user-space applications.
No wonder the usage of eBPF for application and system level profiling/tracing, network level monitoring and security is on the rise \cite{ebpf1,ebpf2}.
Hence, we adopt eBPF for kernel level tracing of the application behavior and to implement security measures that enable to detect and deter the ransomware.
\begin{figure}[h]
\centering
\includegraphics[width=\columnwidth]{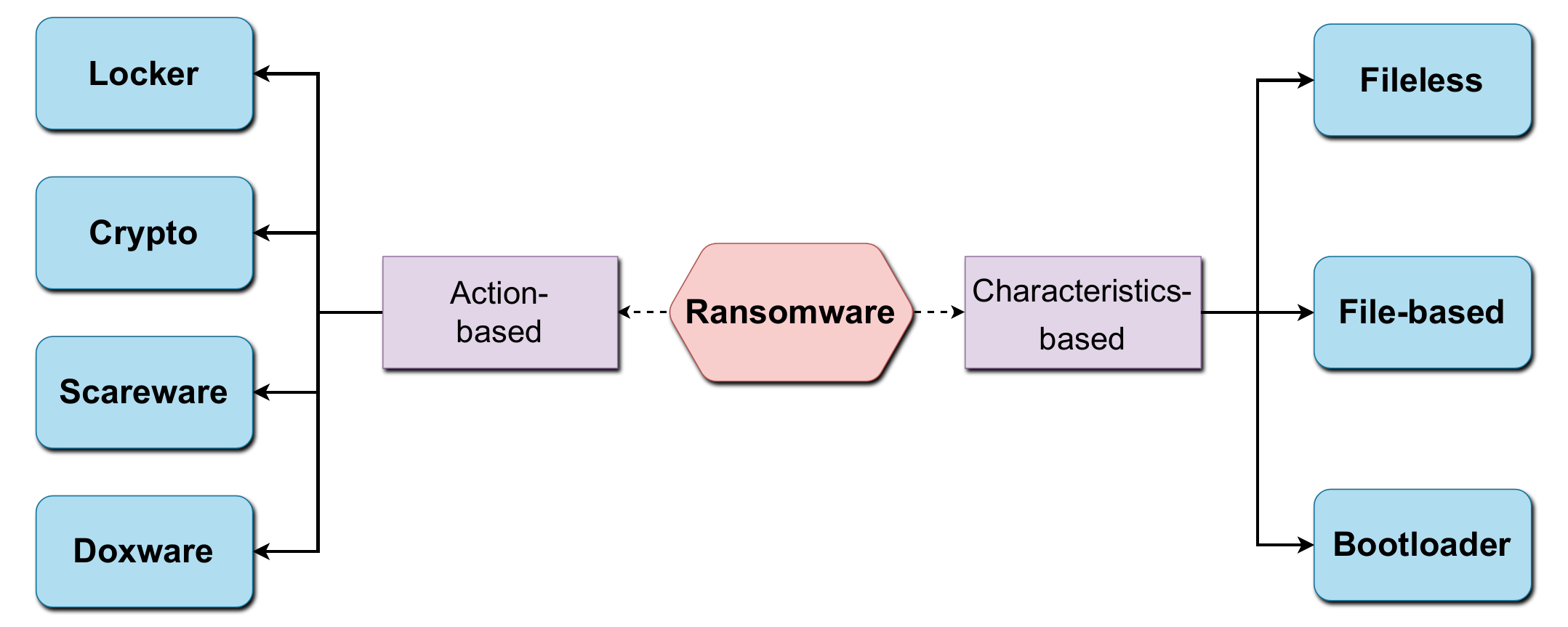}
\captionsetup{justification=centering}
\vspace{-4mm}
\caption{Classification of Ransomware}\label{fig:ransomware-classification}
\vspace{-4mm}
\end{figure}

\sknote{discuss ML approaches and how NLP fits in.}
Natural language processing (NLP) \cite{Bird_Klein_Loper_2009} \textemdash{} a field of artificial intelligence that enables computers to understand and interpret text information. NLP encompasses tasks such as sentiment analysis, spam detection, and topic categorization to automate the organization and understanding of textual data. We develop a NLP 
based semantic analysis model to detect ransomware through analysis of ransom notes.
\noindent The key contributions of our work include:
\vspace{-2mm}
\begin{itemize}
    \item We perform an exhaustive analysis of existing ransomware to identify their distinguishing features. This involves running various ransomware samples in a controlled environment to monitor their system interactions, such as file operations, CPU usage, and system call invocations. By tracing these activities using eBPF, we uncovered specific patterns indicative of ransomware activity. (\S \ref{sec:design}).
    \item We present a novel two-phased approach for ransomware detection and deterrence composed of static and dynamic modules. These modules work together to provide a robust detection framework that operates efficiently at the kernel level. (\S \ref{sec:ps}).
    \item We develop eBPF modules for static (Algo.\ref{alg:static_analysis}) and dynamic analysis (Algo.\ref{alg:dynamic_analysis}). The static analysis module attaches to various exec syscall tracepoints to log the execution of new processes for hash-based comparison against known ransomware samples. The dynamic analysis module monitors and logs file system activities to detect suspiciously high file creation rates typical of ransomware operations. (\S \ref{sec:rtd}).
    \item We develop and train a NLP model 
    to detect ransom notes in real-time (\S \ref{sec:mld}). and demonstrate the efficacy of our proposed solution to detect zero-day 
    attacks in real-time with $\sim{}99.7\%$ accuracy. (\S \ref{sec:eval}).
\end{itemize}


\hide{
\begin{figure*}[ht]
\centering
\includegraphics[width=\linewidth]{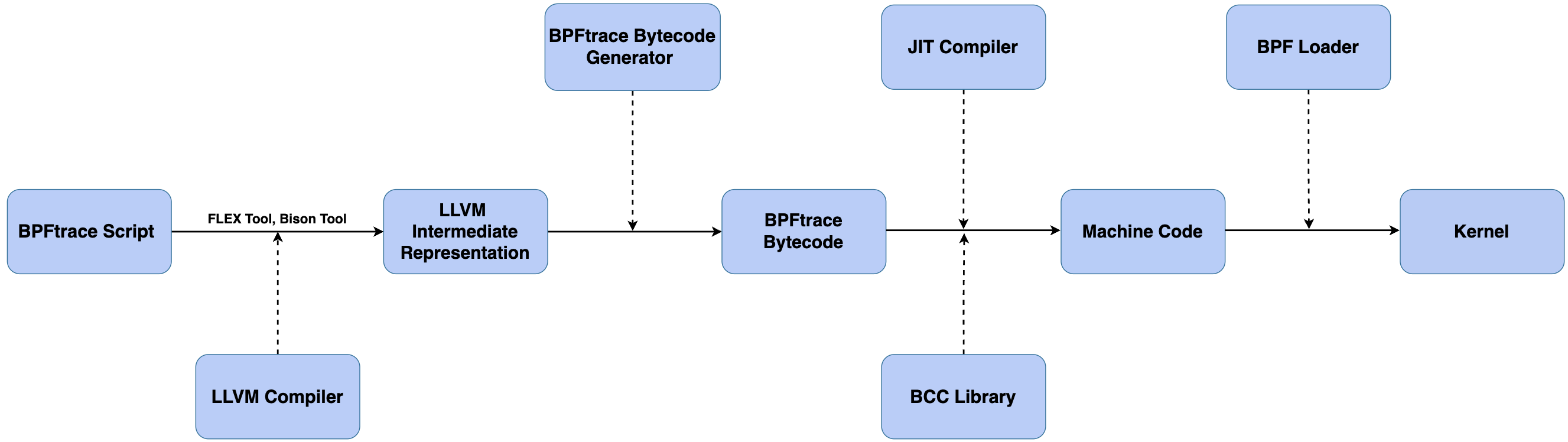}
\captionsetup{justification=centering}
\caption{Flowchart explaining the steps involved while executing an eBPF Program.
\sknote{Not required!}}
\end{figure*}
}


\section{Background}\label{sec:bkgd}
In this section, we present a brief background about classification of ransomwares and eBPF that is relevant to the context of our work.
\subsection{Classification of Ransomware}
Ransomware can be classified based on i) the kind of action it performs (action-based) and ii) the characteristics of its operation (characteristics-based)~\cite{BEAMAN2021102490}.
Figure~\ref{fig:ransomware-classification} shows the two kinds of ransomware classifications.
The actions of ransomware delineate its immediate impact on the victim’s system, ranging from encrypting files and locking down access to outright data exposure threats. Simultaneously, the characteristics of ransomware delve into its underlying attributes, such as the delivery methods, sophistication levels, and overarching strategies employed by cybercriminals. 
\newline\noindent\textbf{Classification based on Action}:
\vspace{-2mm}
\begin{itemize}
    \item \textbf{Locker Ransomware}: This type of malicious software denies users access to their entire systems by locking them out. Instead of encrypting files, it immobilizes the operating system, demanding a ransom to restore regular access.
    \item \textbf{Crypto-Ransomware}: It is a malicious software that encrypts a victim’s files, making them inaccessible to the user. Threat actors demand a ransom for the decryption key needed to unlock the files, often causing significant disruptions and data loss.
    \item \textbf{Scareware}: Scareware is deceptive software that uses alarming messages or fake security alerts to trick users into believing their computer is infected with malware. Its primary goal is to prompt users to purchase unnecessary or fraudulent security products or services.
    \item \textbf{Doxware}: Doxware, also known as leakware, is ransomware that threatens to expose sensitive or private information unless the victim pays a ransom. Instead of encrypting files, it focuses on extorting personal data, potentially causing reputational and privacy damage.
\end{itemize}
\noindent\textbf{Classification based on Characteristics}:
\vspace{-2mm}
\begin{itemize}
    \item \textbf{File-less Ransomware}: It is a type of malicious software that operates without leaving traditional executable files on the victim’s system. It often exploits system vulnerabilities or uses scripts to execute in memory, making detection and prevention more challenging for traditional antivirus tools.
    \item \textbf{File-Based Ransomware}: Traditional file-based ransomware is a form of malicious software that encrypts a victim’s files, rendering them inaccessible. The attackers then demand a ransom payment, typically in cryptocurrency, to provide the decryption key to restore access to the encrypted files.
    \item \textbf{Bootloader Ransomware}: It is a type of malicious software that targets and modifies the computer’s bootloader, disrupting the normal boot process. This ransomware makes the system unbootable by corrupting the bootloader, demanding a ransom to restore standard boot functionality.
\end{itemize}
\subsection{Extended Berkeley Packet Filter \textit{eBPF}}
eBPF is a powerful and resourceful in-kernel virtual machine. 
Although, originally designed for network packet filtering, eBPF has evolved into a universal framework that allows executing custom code snippets in a sandboxed environment within the kernel without the need to recompile the kernel or addition of kernel modules~\cite{calavera2019linux}. 
This flexibility enables dynamic and efficient analysis of various aspects of system behavior, which has lead to widespread adoption of eBPF in areas such as system profiling, tracing, security, and performance optimization.
\hide{
\newline\noindent{\textbf{Why \textit{eBPF}:}}
eBPF offers several advantages over user-space implementation in our approach. Implementing detection entirely in user space would result in higher processing overhead and CPU utilization due to additional context switches, resulting in increased latency for detection, which could slow down the response time. In contrast, eBPF operates directly in the kernel, minimizing overhead and providing real-time monitoring with lower latency. Moreover, eBPF programs are executed in the kernel, making it harder for ransomware to disrupt than user space processes.
}

\section{Design Rationale}\label{sec:design}
Traditional analysis techniques often overlook critical system interactions and nuances that could provide valuable insights into the complex mechanisms employed by modern ransomware variants.
Hence, we first identify the key operational characteristics and run-time behavior of file based crypto ransomwares. 
We run an ensemble of ransomwares including REvil, HelloKitty, IceFire and WannaCry on a test virtual machine to unearth the following distinct, yet common features. Table~\ref{tab:rsfeatures} summarizes our key findings. We discuss below the following observations. 

\noindent\textbf{eBPF based system call tracing:}
Here, we specifically instrument the Linux kernel with eBPF 
to monitor all of the 336 tracepoints available in \textsl{sys\textunderscore enter} system call events. 
This, not only provides an unprecedented level of visibility into the granular activities occurring during a ransomware execution\footnote{This exhaustive tracing is possible in our test environment where only the ransomware application is run.}, but also offers a more reliable and consistent mechanism than the kernel probes regardless of the variability in kernel versions. 
We uncovered several distinctive features and behaviors 
prevalent during ransomware operations but relatively uncommon in benign system activities. 
Such key features include i) high CPU usage, ii) frequent invocation of system calls such as \textsl{ openat, unlinkat, pkill, rename, write,} \textsl{OpenSSL calls}, and iii) extensive use of \textsl{urandom} file. 
\newline\noindent\textbf{\textsl{urandom}}: Regardless of the specific encryption algorithm or library used, any encryption process requires a source of random numbers for securely generating encryption keys. Our analysis revealed that sophisticated advanced ransomware variants like REvil heavily utilize the `/dev/urandom' pseudo-random number generator to aid their encryption processes. This distinctive behavior left a characteristic trace that could not have been identified by merely tracing the OpenSSL or other existing encryption libraries. 

\begin{table}[t]
\caption{Primary Syscall Invocations Across Ransomware Variants During Ransomware Execution}\label{tab:rsfeatures}
\scriptsize
\begin{tabular}{cccc}
\hline
\scriptsize
\textbf{Feature} & \textbf{REvil} & \textbf{HelloKitty} & \textbf{WannaCry} \\
\hline
Unlink usage & High & Med & High \\
Urandom utilization & High & Low & Low \\
Pkill invocation & High & High & High \\
Rename syscall & High & High & High \\
Write syscall & High & High & High \\
OpenSSL usage & Low & High & High \\
CPU utilization & High & High & High \\
Openat usage & High & High & High \\
\hline
\end{tabular}
\end{table}
\noindent\textbf{\textsl{Encryption}}: Further, we observed that ransomware variants like REvil do not rely on standard encryption libraries like OpenSSL or mbedTLS, which can be traced by placing uprobes on libcrypto.so or libmbedtls.so respectively. Instead, 
they employ custom-loaded encryption libraries with novel encryption techniques such as the SALSA-20 cipher observed in REvil.
It is important to note that while sophisticated ransomware like REvil employs custom encryption techniques and avoids standard libraries, some more primitive variants like WannaCry and Hellokitty ransomware directly leverage OpenSSL for encryption. In such cases, placing uprobes on the dynamic symbols of OpenSSL library (libcrypto.so) i.e. \textit{EVP\_EncryptInit\_ex} and \textit{EVP\_CipherInit\_ex},  effectively capture their encryption activities.
\newline\noindent\textbf{\textsl{File Operations}}:
The reason for high unlink/unlinkat system call usage is because the ransomware variants often create temporary .lck or .temp files as intermediary files between the original file and the final encrypted file. After creating the final encrypted file, the ransomware deletes these temporary files using the unlink or unlinkat system calls.
The high frequency of pkill system call invocation is observed because ransomware typically uses pkill to terminate all running processes before encrypting them. Additionally, the command pkill kills open file handles, ensuring no interruptions occur during the file encryption process.
The renaming of files from their original names to custom encrypted file extensions leads to the frequent invocation of the \textit{sys\_enter\_rename} system call. This behavior is a characteristic of ransomware operations, as it renames the encrypted files with specific extensions to indicate their encrypted state.
\begin{figure}[t]
\centering
\includegraphics[width=\columnwidth]{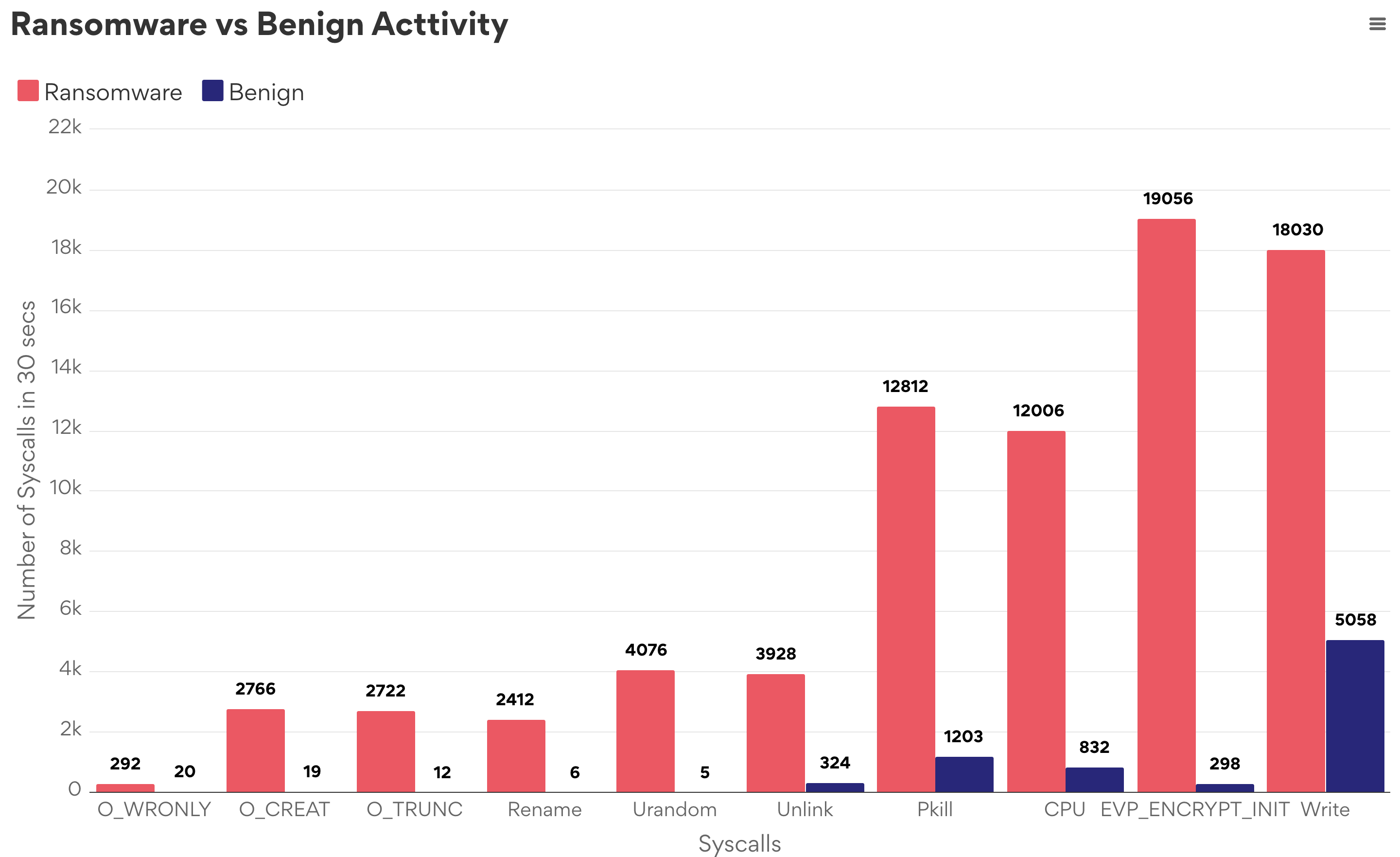}
\captionsetup{justification=centering}
\caption{Analysis of Syscall Invocation Frequency: Ransomware vs. Benign Processes}
\label{figure:ransom}
\end{figure}
\sknote{Need to replace this with BAR plot.}
By comprehensively tracing all 336 \textit{sys\_enter} system calls, we uncovered previously unseen aspects of the ransomware attack process, revealing the intricate mechanisms and operations employed by these sophisticated malware variants. 



The graph in Fig.~\ref{figure:ransom} illustrates the comparative frequency of system call invocations observed during ransomware execution versus benign processes. The data for ransomware processes was collected by running various ransomware samples for approximately 30 seconds while encrypting 1.5GB of data. The data for benign processes was obtained from normal system usage over a 30-second period. 


\section{Proposed Solution}\label{sec:ps}
\begin{figure}[t]
\centering
\includegraphics[width=\columnwidth]{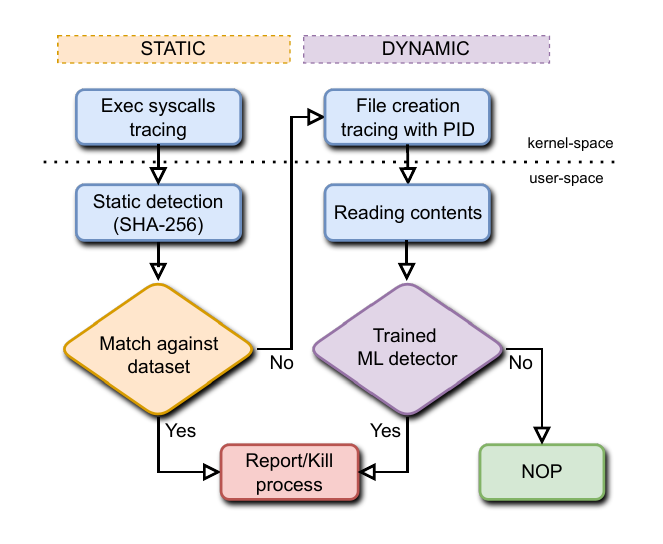}
\captionsetup{justification=centering}
\caption{Ransomware Detection Pipeline}\label{fig:2pdesign}
\end{figure}

We propose a novel two-phased approach that combines static and dynamic behavioral analysis techniques.  The key dynamic property we focus on is to monitor the creation of ransom notes and leverage machine learning to classify the ransom notes as malicious or benign. Fig.~\ref{fig:2pdesign} shows the proposed two-phase ransomware detection pipeline.














\noindent\textbf{Static Analysis}: Here, we use eBPF to place a tracepoint on the exec system calls to capture new processes being executed. We then perform hash-based detection on the process by comparing its hash with a dataset obtained from MalwareBazaar \cite{MalwareBazaar}, which consists of SHA-256 hashes of about 777,073 malware samples. This static analysis layer serves as the initial line of defense.
\newline\noindent\textbf{Dynamic Analysis}:
When the static analysis layer fails to detect any malicious indicators, i.e. zero-day attacks, the dynamic (second) layer of our approach takes over, focusing on the behavior-based detection of ransom note creation, a distinctive and essential step in the ransomware attack lifecycle. 
Ransomware typically follows a well-defined sequence of actions, including file encryption, followed by the creation and delivery of a ransom note containing instructions for payment and file recovery. While existing behavior-based techniques may monitor file system activities or system call patterns, our technique aims to provide a more targeted and effective defense against ransomware operations by detecting and responding to this critical step in the ransomware kill chain. After analyzing the system calls invoked during ransomware operations, we focused on observing and analyzing the file creation patterns using the openat system call. The tracepoint at \textit{sys\_enter\_openat} has an argument (flags) of type int, which can be decoded using the definitions in /usr/include/asm-generic/fcntl.h. By understanding these flag values, we can determine the file access modes being used. For instance, the flag 0x00000100 corresponds to O\_CREAT, indicating a "create-only" file access mode. O\_RDONLY (0x00000000) represents read-only access, and so on. While observing the file creation patterns and correlating them with the fcntl.h definitions, we noticed a crucial pattern: ransom notes were consistently created with the O\_CREAT flag across various ransomware variants. This observation led to our approach of monitoring the openat system call with the O\_CREAT file access flag to detect the creation of new files, which the generation of a ransom note would trigger during a ransomware operation. When a new file is created, its content is analyzed using machine learning models trained to distinguish between genuine ransom notes and benign files. This classification is based on various features extracted from the file, such as textual patterns, keywords, formatting, and other relevant characteristics.

If the created file is classified as a potential ransom note, our system can take immediate action, such as terminating the responsible process or quarantining the affected system to prevent further damage. By targeting this critical stage of the ransomware attack chain, our approach aims to disrupt the ransomware operation before file encryption and extortion demands can be completed. It is crucial to note that the effectiveness of our technique in protecting files varies depending on the specific ransomware behavior. The majority of crypto-ransomware variants primarily fall into two categories: those that create a ransom note after encrypting each directory and those that create a ransom note before encrypting any files. In the former case, our technique may sacrifice a few files and folders before detecting and disrupting the ransomware operation. However, it can prevent the encryption of other substantial amounts of files and folders. In the latter case, where the ransom note is created before any encryption occurs, our technique can protect all files from being encrypted.

Sophisticated ransomware variants often stay idle for extended periods, slowly encrypting small volumes of files without drawing attention. In such cases, detection techniques relying solely on high CPU usage, unlink/write activity, or high I/O operations may not yield impressive results, as these variants maintain a low overhead. Therefore, behavioral analysis focused on detecting ransom note creation after the encryption of a directory becomes an important aspect of identifying these sophisticated ransomware variants.

\begin{algorithm}[t]
\caption{ \textit{eBPF} Pseudocode for Static Analysis}
\label{alg:static_analysis}
\begin{algorithmic}[1]
\footnotesize
\STATE // Attach to tracepoints for exec\{v,l,ve,cl,lp,vp\} syscalls
\STATE \textbf{Function} \textsc{static\_analysis}()
\STATE \textbf{On} sys\_enter\_execve trigger:
\STATE \quad pid = get\_current\_process\_id();
\STATE \quad comm = get\_current\_process\_name();
\STATE \quad log\_execution(pid, comm);
\STATE \textbf{On} sys\_enter\_execv trigger:
\STATE \quad pid = get\_current\_process\_id();
\STATE \quad comm = get\_current\_process\_name();
\STATE \quad log\_execution(pid, comm);
\STATE \algorithmicreturn \sknote{???} \newline
\STATE \textbf{End Function}
\end{algorithmic}
\end{algorithm}

\subsection{Real-Time Detection}\label{sec:rtd}

The real-time ransomware detection system integrates eBPF for dual-layer security by combining static and behavioral analysis. Initially, eBPF tracepoints are placed on the exec syscalls to track program executions as depicted in Algo.~\ref{alg:static_analysis}. A Python script computes the SHA-256 hash of each executable and checks it against a malware hash dataset, terminating any malicious process immediately. Concurrently, the same eBPF program also monitors file creation events, capturing the file path of newly created files if the number of creation events per process exceeds a threshold value T, with the flag O\_CREAT attribute and the corresponding Process ID (PID) of the process initiating the creation, i.e. ransomware in the case of ransom note creation (lines 18,19 Algo.~\ref{alg:dynamic_analysis}). A Python script then retrieves these file contents and uses our pre-trained machine-learning model to classify them as potential ransomware ransom notes or benign files. Upon detecting a ransom note, the script swiftly terminates the associated process using the PID captured earlier. This swift response disrupts ransomware operations within moments of the ransom note's creation, preventing further file encryption attempts. As illustrated in Fig.~\ref{fig:2pdesign}, this integrated system architecture enables rapid and automated detection and mitigation of ransomware attacks.



\subsection{ML-based Detector}\label{sec:mld}


\begin{algorithm}[t]
\caption{ \textit{eBPF} Pseudocode for Dynamic Analysis}
\label{alg:dynamic_analysis}
\begin{algorithmic}[1]
\footnotesize
\STATE \textbf{Function} \textsc{dynamic\_analysis}()
\STATE \#include $<$asm-generic/fcntl.h$>$
\STATE \#define FILE\_CREATION\_THRESHOLD T
\STATE config = \{
\STATE \quad stack\_mode = "perf",
\STATE \quad max\_strlen = 128
\STATE \};
\STATE \textbf{On} sys\_enter\_openat trigger:
\STATE \quad uid = get\_current\_user\_id();
\STATE \quad comm = get\_current\_process\_name();
\STATE \quad flags = get\_syscall\_flags();
\STATE \quad \textbf{if} !(pid \textbf{in} process\_counter\_map) \textbf{then}:
\STATE \quad \quad process\_counter\_map[pid] = \{0\};
\STATE \quad \textbf{if} flags \& O\_CREAT \textbf{then}:
\STATE \quad \quad pid = get\_current\_process\_id();
\STATE \quad \quad filename = get\_syscall\_argument(args->filename);
\STATE \quad \quad process\_counter\_map[pid].file\_creation\_count += 1;
\STATE \quad \quad \textbf{if} process\_counter\_map[pid].count $\geq$ T \textbf{then}:
\STATE \quad \quad \quad log\_file\_creation(pid, filename);
\STATE \textbf{End Function}
\end{algorithmic}
\end{algorithm}

We prepare a diverse dataset, comprising both real-world ransomware ransom notes and a wide variety of benign text files. Using this dataset, we train a machine-learning model to accurately distinguish between malicious ransom notes and legitimate files based on their content and characteristics.

To represent the "ransomware" class, we leveraged the publicly available "ransomware\_notes" dataset on GitHub by Threat\_labz \cite{threatlabz_ransomware_notes}. From this repository, we collected 170 unique ransom notes from various ransomware variants. Additionally, we executed seven different ransomware samples in a controlled virtual environment to obtain seven more ransom notes, bringing the total number of unique ransom notes in our dataset to 177.
To represent the "benign" class, we utilized a diverse dataset compiled by Ken Lang \cite{Lang95}, consisting of text files covering a wide range of topics, including technology, sports, politics, business, entertainment, graphics, medicine, space, and more. We collected 177 such files and complemented this dataset with 50 README files collected from various public repositories and code scripts (Python, Shell, C, HTML, etc.) to enrich the benign data diversity.

\begin{figure}[t]
\centering
\includegraphics[width=\columnwidth]{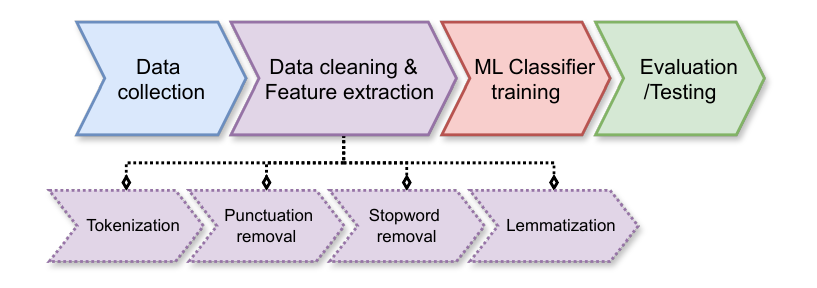}
\captionsetup{justification=centering}
\caption{Machine Learning Pipeline}
\label{fig:mlp}
\end{figure}

We then employ the Natural Language Toolkit (NLTK) library \cite{bird2009natural}, to apply a set of effective pre-processing techniques on the collected text data as depicted in Fig.~\ref{fig:mlp}. 
These techniques include tokenization, punctuation removal, stop word removal and word normalization.
At this stage, it is crucial to preserve the keywords that distinguish ransomware notes from benign text for which we use lemmatization. 

After pre-processing, the lemmatized tokens are joined into a single string, representing the processed content for each sample in the dataset.
To extract meaningful features from the pre-processed text, we employ two techniques, namely, Term Frequency-Inverse Document Frequency (TF-IDF) and Chi-Squared Feature Selection. The TF-IDF approach creates a vector representation of the text, where each dimension corresponds to a word in the corpus, and the values represent the importance of that word in the given text based on its frequency and inverse document frequency. Complementing this, the Chi-Squared Feature Selection method selects the most informative features from the TF-IDF representation based on their chi-squared statistic, which measures the dependence between the features and the target classes (ransomware or benign). We find that using the top 400 features yields the best results in terms of accuracy.
The extracted features are then used to train a Multinomial Naive Bayes classifier, with a train-data to test-data ratio of 7:3 . The trained model, TF-IDF vectorizer, and feature selector are saved to a pickle file for future use in detecting and classifying potential ransom notes.


\section{Preliminary Evaluation}\label{sec:eval}

\textbf{Offline testing on collected data:} Our ML-based detector's remarkable accuracy and effectiveness in detecting ransomware is summarized in table \ref{table:eval}. The model achieved an impressive accuracy of 99.76\% on test data, meaning it could reliably distinguish ransomware ransom notes from benign files. The recall score of 0.9943, is a testament to the model's ability to identify nearly 99.43\% of the actual ransomware ransom notes present. This high recall rate gives us confidence that the model can detect the vast majority of ransomware attacks, minimizing the chances of missing potential threats. Moreover, the F1-score of 99.7\% highlights the model's achievement of both high precision and high recall, which are highly desirable traits for a robust ransomware detection system. Notably, the average cross-validation score of 99.02\%, obtained using a 10-fold cross-validation technique, suggests the model's generalization capability to unseen data. A high cross-validation score indicates a low likelihood of overfitting, as the model's performance is consistently evaluated on multiple hold-out subsets of the data. This high score gives us confidence that the model can maintain its performance when deployed in real-world scenarios without overfitting to the training set.
\newline\noindent\textbf{Real-time testing on unseen ransomware:} In order to validate the efficiency and robustness of our real-time detection technique, we tested the behavior-based component of our detection technique with unseen ransomware samples whose ransom notes were not part of the training data. Our detection model was able to successfully detect the ransomware operation and kill the ransomware process in 
just 5.4 seconds for the IceFire ransomware and in 2.5 seconds for the REvil ransomware as shown in Table~\ref{tab:ransomware_results}. Notably, these ransomware samples typically take about 1.5 minutes to completely encrypt the sample files in the VM if not intervened upon. This real-world testing demonstrates the effectiveness and rapid response capabilities of our solution, minimizing the potential damage caused by ransomware attacks.



\begin{table}[t]
\caption{NLP Test results}
\centering
\begin{tabular}{cc}
\hline
\textbf{Metric} & \textbf{Value} \\
\hline
Accuracy & 0.9976 \\
Precision & 1.0 \\
Recall & 0.9944 \\
F1-Score & 0.9972 \\
Average CV Score (K=10) & 0.9902 \\
\hline
\end{tabular}
\label{table:eval}
\end{table}

\begin{table}[h]
\centering
\caption{Real-Time Detection Results}
\scriptsize
\begin{tabular}{cccc}
\hline
\textbf{Ransomware} & \textbf{Result} & \textbf{Time to Detect (secs)} & \textbf{Affected Files} \\
\hline
HelloKitty & Success & 1.023 & 0.07\% \\
REvil & Success & 2.529 & 0.39\% \\
IceFire & Success & 5.403 & 0.56\% \\
Kuiper & (FN) & - & 100\% \\
\hline
\end{tabular}
\label{tab:ransomware_results}
\end{table}


\section{Related Works} \label{sec:rw}
Existing ransomware detection techniques primarily fall into two categories: Static  and Dynamic approaches. While these methods provide valuable insights into ransomware activities, they often exhibit several limitations such as failure to detect zero-day attacks, low accuracy, higher computational overhead 
and longer detection times. 
\newline\noindent{\textbf{Static detection techniques}}
employ hash-based and/or signature-based detection which rely on predefined patterns to identify ransomware executable and/or their variants. These patterns are typically derived from known ransomware samples. 
While static analysis can effectively detect known ransomware strains with high accuracy, it struggles with detecting polymorphic or previously unseen variants, \textit{i.e.}, zero-day attacks \cite{4413008_limitations}. Moreover, adversaries can easily evade detection by altering the code structure or through the obfuscation of ransomware executable, rendering static-based methods less effective in dynamic threat landscapes. 
Zhang \textit{et al.}~\cite{ZHANG2019211} proposed a static analysis framework based on N-gram opcodes with machine learning for ransomware classification. 
However, handling obfuscated or polymorphic code still remains an open concern.
\newline\noindent{\textbf{Dynamic detection techniques}}
involve runtime monitoring and analysis of system activities and behaviors 
to detect anomalies, suspicious patterns, or deviations from established norms that may indicate the presence of ransomware or other malicious code. 
By analyzing runtime behavior, these methods can potentially identify previously unknown or emerging ransomware variants, including zero-day attacks. 
However, behavior-based approaches may generate false positives when the legitimate applications exhibit similar behavior patterns, resulting in reduced accuracy and increased operational overhead. 
Kharraz \textit{et. al.}~\cite{kharraz} proposed monitoring abnormal file system activities as a mechanism to identify ransomware. They further extended their research by developing UNVEIL~\cite{unveil}, a dynamic analysis system that detects ransomware by monitoring and analyzing the behavioral patterns of file system activities such as suspicious I/O operations, file modifications, and changes to the desktop environment that are indicative of ransomware activities.


\hide{in our approach we use a two-phased detection technique which consists of hash-based detection and behavioral-based detection that focuses on detecting creation of ransom notes as described in Section \ref{sec:ps}}

Scaife \textit{et al.}~\cite{scaife} developed CryptoDrop, employing techniques such as calculating file entropy using Shannon’s formula, measuring file version similarity through hash comparisons, and tracking file deletions to detect ransomware. Similarly, Chen \textit{et al.}~\cite{Chen2018TowardsRM} proposed monitoring file deletions, creations, renames, and changes for ransomware detection. Our behavioral model augments these work by detecting the creation of ransom notes.

\sknote{Need to cite: \cite{higuchi2023real,zhuravchak2023monitoring}}

\section{Conclusion and Future work}\label{sec:conclusion}

To summarize, we proposed a two-phase approach for timely detection and deterrence of ransomware attacks by developing new eBPF and ML algorithms. The novelty of our work lies in the adoption of eBPF for automated static analysis that gets triggered at the time of execution of new processes in the system. Further, we have also presented a novel behavior-based ransomware detection using the eBPF and NLP through the detection of ransom notes. 
\newline\noindent\textbf{Future Work}: We are working on extending the static analysis phase with the creation of pattern matching based Yara rules that can further improve the detection capabilities. We also aim to develop a more robust behavior model that can build on the sequence distinct process/file operations and correlate the time-series data to aid for robust recurrent neural network based models to detect any zero-day attacks.

\bibliographystyle{ACM-Reference-Format}
\bibliography{reference}

\end{document}